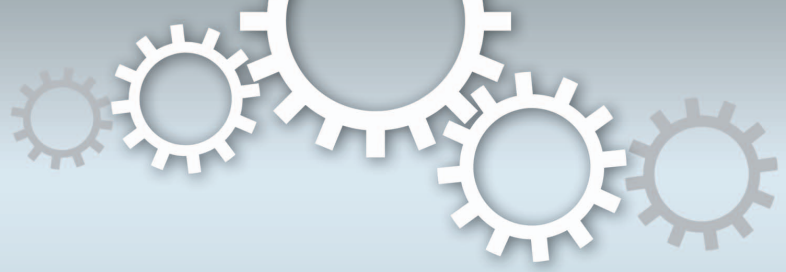

**OPEN**

# Controlling Networks of Nonlinearly-Coupled Nodes using Response Surfaces

Jason Shulman[1], Franck Malatino[1], Alexander Mo[2], Killian Ryan[3] & Gemunu H. Gunaratne[3]

[1]Department of Physics, Richard Stockton College of New Jersey, Galloway, NJ 08205, [2]Bellaire High School, Bellaire, TX 77041, [3]Department of Physics, University of Houston, Houston, TX 77204.

Correspondence and requests for materials should be addressed to G.H.G. (gemunu@uh.edu)

Control of complex processes is a major goal of network analyses. Most approaches to control nonlinearly coupled systems require the network topology and/or network dynamics. Unfortunately, neither the full set of participating nodes nor the network topology is known for many important systems. On the other hand, system responses to perturbations are often easily measured. We show how the collection of such responses –a response surface– can be used for network control. Analyses of model systems show that response surfaces are smooth and hence can be approximated using low order polynomials. Importantly, these approximations are largely insensitive to stochastic fluctuations in data or measurement errors. They can be used to compute how a small set of nodes need to be altered in order to direct the network close to a pre-specified target state. These ideas, illustrated on a nonlinear electrical circuit, can prove useful in many contexts including in reprogramming cellular states.

Natural and artificial systems such as those underlying biological processes[1,2] and the internet[3] consist of collections of interacting elements, and can be modeled as networks of coupled nodes[1,4]. Feedback within such networks is believed to buffer their response under external changes and minor alterations in the systems[5]. The feedback-induced moderation lies at the heart of robustness, for example, of biological processes. Unfortunately, the robustness also makes it difficult to implement controlled alterations of the underlying system. Thus, modifying one or a few nodes may be countered by feedback or may initiate uncontrolled, and possibly damaging, consequences. The surprising lack of efficacy of many drugs designed to act on single molecular targets[6,7] and harmful side-effect from medications (*e.g.*, Vioxx[8]) are a testament to these difficulties. Such problems can be addressed if techniques to direct nonlinearly coupled networks to target states were available.

Precise controllability, one of the major goals of network analyses, has inspired several recent studies. *Controllability* of a system is defined as the ability to steer the system from any initial state to an arbitrary final state in finite time[9,10]. In networks, the number of *control nodes* depends primarily on the degree distribution but not on the clustering coefficient or the community structure of the network[11,12]. Furthermore, contrary to expectations, it was found that the driver nodes are typically not the high-degree nodes of a network[11]. However, *scale-free networks* can be driven to a pre-specified state[13] or steered to a desired evolution[14] by pinning or targeting their most highly-connected nodes. Ref. 15 analyzes graph structures of controllable linear systems[16], and shows that nodes required for network control include source nodes (*i.e.*, those with only outgoing links), internal dilations (due to branching points), and external dilations (due to surplus nodes with only incoming links). The application of these algorithms requires knowledge of the network topology. In contrast, the approach introduced in Ref. 17 (which applies to linearly or nonlinearly coupled networks) relies on perturbation of control nodes and following the evolution of the system to search for passage from an undesirable initial state to the basin boundary of a desirable target state. It was used to identify possible therapeutic targets for T-LGL leukemia and to drive an associative memory network to a target state[17].

Unfortunately, the network topology of many real systems is not known[18], and it is difficult to follow the time evolution of other systems (*e.g.*, gene regulatory networks) experimentally. Here, we propose a methodology to address a more restricted issue on partially-known systems, namely how to move such a system close to a *pre-specified* target state. Although the methodology applies for a broad class of systems, we introduce terminology from gene regulatory networks for ease of presentation. The *state of the system* is the collection of values taken by the nodes. The state is to be changed by altering the values of a small set $n$ of nodes, defined as the *master nodes*; the remaining nodes are *slave nodes*. States obtained by externally setting the values of the master nodes are referred to as *mutants*. A *single knockout mutant* is a state where the value of one of the master nodes is set to zero and a *double knockout mutant* is one where two master nodes are set to zero. The object of our study is the *response surface* containing the states of all master node mutants[19,20]. It is an *n*-dimensional surface in state space. The nature of the response surface depends on network interactions; conversely, the response surface itself contains partial information about network interactions.

Figure 1 outlines the method schematically. A network of the type shown in Figure 1(a) may represent a system under study. When neither its topology nor the forms of interactions between its nodes are known, it is not possible to use the network for control. Alternatively, we can represent the behavior of the system in state space. As an example, consider a network to be controlled using a single master node, whose value is denoted $X_1$. As $X_1$ is externally manipulated, the state of the system spans a one-dimensional curve $\gamma$, which in principle can be accessed experimentally. (For example, the response of a genetic regulatory network to changes in one biomolecule can be measured using microarrays or deep sequencing[21].) The solid line in Figure 1(b) shows the cross-section of $\gamma$ in the $(X_1, X_7)$ plane. $\mathcal{P}_0$ and $\mathcal{P}_1$ represent the states of the original system and the single knockout mutant, and $\mathcal{T}$ denotes the desired target state.

The first step is to find a sufficiently accurate approximation to $\gamma$ in the region of interest. The lowest order (linear) approximation can be made using $\mathcal{P}_0$ and $\mathcal{P}_1$; higher order approximations require additional points on $\gamma$. Next, compute the point $t$ on the approximation that is closest to $\mathcal{T}$. By setting $X_1$ to its value at $t$ externally, the system is forced to a point on the response surface in close proximity to $\mathcal{T}$. This is the best that can be done by only manipulating $X_1$. However, since $\gamma$ is a one-dimensional curve in a high dimensional state space, $t$ may not be sufficiently close to $\mathcal{T}$. Thus, we need to expand the set of master nodes. Interestingly, as shown below, the states $\mathcal{P}_0$, $\mathcal{P}_1$, and $\mathcal{T}$ can be used to identify the "best" node to be selected as the second master node; in the example, it is assumed to be node 3. Next, we consider the response surface for the pair of master nodes shown in Figure 1(c). As before, it is approximated using the plane $\mathcal{P}_0\mathcal{P}_1\mathcal{P}_3$, the last point being the state of the single knockout mutant of node 3. Now, the point closest to $\mathcal{T}$ on $\mathcal{P}_0\mathcal{P}_1\mathcal{P}_3$, and the next node to be included in the master set are computed; the process is continued until a point sufficiently close to $\mathcal{T}$ is reached. Figure 1(d) shows how the distance to $\mathcal{T}$ reduces with the number of master nodes in our example.

*The novel aspect of our approach is its reliance only on the response surfaces; it does not require or suppose additional information about the network.* An important observation from studies of model networks is that these response surfaces are smooth[20] and hence can be well-approximated using low-order polynomials. Computations of the coarsest approximations, *i.e.*, planes, only require responses of the single knockout mutants[19], which were utilized in the example outlined in Figure 1. Finally, stochasticity in the measurements is not

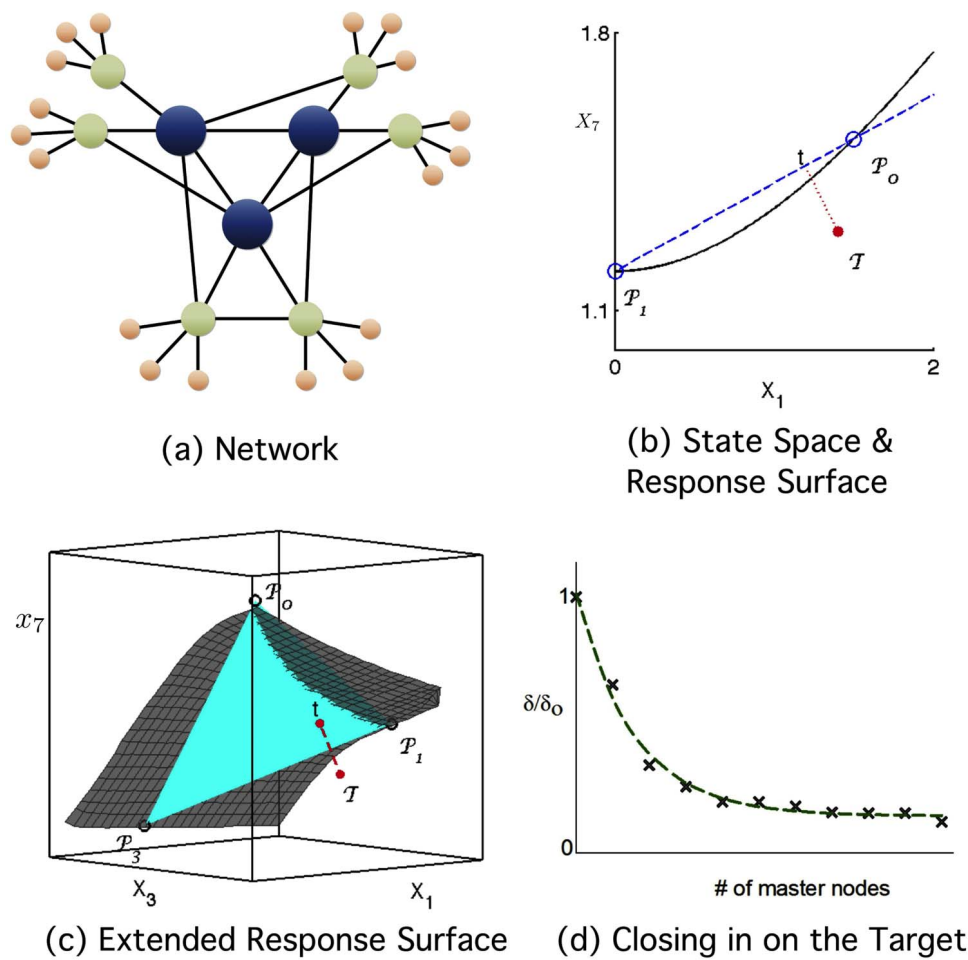

(a) Network

(b) State Space & Response Surface

(c) Extended Response Surface

(d) Closing in on the Target

**Figure 1** | (a) A schematic of a network. When its topology and the forms of interactions between its nodes are not known, the network cannot be used to move the system to a target state. (b) The response surface and a linear approximation when the single master node 1 is externally altered. $\mathcal{P}_0$, $\mathcal{P}_1$, and $\mathcal{T}$ denote the states of the original system, the single knockout mutant, and the target state. The closest point t on the approximation and the next master node can be computed from $\mathcal{P}_0$, $\mathcal{P}_1$, and $\mathcal{T}$. (c) The response surface and its planar approximation when there are two master nodes 1 and 3. (d) The decay of the closest distance $\delta$ as a function of the number of master nodes. $\delta_0$ is the distance between the states of the initial system and the target.

amplified in computations of the response surfaces. Thus, the proposed algorithm provides a realistic approach for controlling a partially known network to a target state.

## Methods

The control algorithm is illustrated using a nonlinearly coupled electrical circuit. Nonlinear elements of the circuits are junction field-effect transistors (JFETs)[22], semiconductor devices containing a source (S), a drain (D) and a gate (G). Conduction between S and D is modulated by the gate voltage; the current $I$ depends on the potential differences $V_{DS}$ between D and S and $V_{GS}$ between G and S. We have shorted G and S (i.e., $V_{GS} = 0$) reducing the number of terminals in each JFET to two. Consequently, the polarity independence of S and D conduction is broken. For example, an $n$-channel device exhibits $p - n$ junction diode conduction for $V_{DS} < 0$ and typical JFET current-voltage characteristics for $V_{DS} > 0$; the latter is shown in Figure 2(c).

The circuit we analyze is shown in Figure 2(a), which is equivalent to the network of Figure 2(b). It has two regulatory levels. Each node of the circuit is a point of equipotential. Interactions between nodes are determined by resistors/JFETs connecting them. Nodes 1, 2, 3, and 4 on the uppermost regulatory level are selected as (preliminary) master nodes. This circuit only contains activating interactions. Below and in Supplementary Materials, we summarize corresponding results from a network containing "inhibitory" interactions as well.

The circuit was driven at 10 V by a Topward 6302D (TekNet Electronics, Alpharetta, GA.) power supply. The 16 node voltages were measured with two 8-channel AD cards (MiniLab-1008, Measurement Computing Corp.) and in a few cases cross checked using hand-held voltmeters. Mutants are generated by setting the master node potentials externally using either the AD cards or by connecting to power supplies. Single (double) knockout mutants are obtained by grounding one (two) nodes. Since the master nodes themselves are coupled, controlling them at or close to the required potentials is a non-trivial task. Control was particularly difficult when the power supplies were sourcing small voltages, since they have to accept current from the network rather than supply current to it. Another difficulty is the interference between power supplies due to their small but non-zero internal impedance. These difficulties were resolved by placing a small resistor (typically 5Ω) in parallel with the power supply controlling each master node. It allowed the supply to source current, most of which passed through the resistor, and to apply the desired voltage on the node. Implementing simultaneous controls on multiple nodes was feasible with this approach.

The state of the network (i.e., the set of node potentials) varies smoothly as master node potentials are altered. In Figure 3, this is illustrated through a cross section $X_7(X_1, X_3)$, where $X_n$ is the $n^{th}$ node potential. The coarsest approximation to the surface is a plane computed as follows: denote the state of the unperturbed circuit by $\mathbf{X}^{(0)} \in \mathbb{R}^{16}$, and those of the four single knockout mutants by $\mathbf{X}^{(n)} \in \mathbb{R}^{16}$, for $n = 1, 2, 3, 4$. The 4-dimensional plane in $\mathbb{R}^{16}$ passing through these points can be parameterized as

$$\mathbf{X}(\lambda) = \mathbf{X}^{(0)} + \sum_{n=1}^{4} \lambda_n \left( \mathbf{X}^{(n)} - \mathbf{X}^{(0)} \right). \tag{1}$$

Next, it is necessary to model interactions between master nodes. Denoting the projections of $\mathbf{X}^{(n)}$ ($n = 0, 1, 2, 3, 4$) to the subspace of master variables by $\mathbf{x}^{(n)}$, linear approximations to these interactions can be expressed as

$$\mathbf{x}^{(n)} - \mathbf{x}^{(0)} = \sum_{m \neq n} \mu_{nm} \left( \mathbf{x}^{(m)} - \mathbf{x}^{(0)} \right), \tag{2}$$

for $m, n = 1, 2, 3, 4$. These equations can be written in terms of a $4 \times 4$ matrix $\mathbf{M}$ which satisfies $\mathbf{M} \cdot (\mathbf{x}^{(n)} - \mathbf{x}^{(0)}) = 0$ for each $n$. $\mathbf{M}$, with unit diagonal elements, can be computed from the data $\mathbf{x}^{(0)}$ and $\mathbf{x}^{(n)}$ [19]. $\mathbf{M}$ approximates the motion of the state on the response surface as one or more master nodes are altered.

The approximation to a response surface can be improved using higher order polynomials. For example, the quadratic approximation to the response surface can be expressed as

$$X_i\left(v^{(1)}, v^{(2)}\right) = X_i^{(0)} + \sum_{n=1}^{4} v_{in}^{(1)} \left(X_n - X_n^{(0)}\right) \\ + \sum_{n \leq m} v_{imn}^{(2)} \left(X_m - X_m^{(0)}\right)\left(X_n - X_n^{(0)}\right) \tag{3}$$

where $i$ is a slave node, and $m, n$ are master nodes. The coefficients $v_{in}^{(1)}$ and $v_{imn}^{(2)}$ are computed by forcing the approximation to pass through a set of mutant states. This system has to be supplemented by quadratic generalizations of Eq. (2). The solution of the quadratic approximation requires data on $\frac{1}{2}d(d+1)$ additional mutants, $d$ being the dimensionality of the response surface.

Once an approximation to the response surface is evaluated, we can compute how the master nodes need to be altered in order to reach as close as possible to a target state $\mathbf{T} \in \mathbb{R}^{16}$. We illustrate the computation using the planar approximation, Eq. (1). Note that changes in the master nodes only move the system along the response surface, which is approximated by $\mathbf{X}(\lambda)$. The strategy is to search for $\mathbf{X}\left(\tilde{\lambda}\right)$ that minimizes the weighted-square-distance $\sum w_m (T_m - X_m(\lambda))^2$. The weights $w_m$ account for cases where the proximity of certain nodes to $\mathbf{T}$ are more critical. The minimum $U(\mathbf{T}; \mathbf{w})$ satisfies

$$\sum_{m=1}^{16} w_m \left(T_m - X_m\left(\tilde{\lambda}\right)\right)\left(X_m^{(n)} - X_m^{(0)}\right) = 0, \tag{4}$$

for $n = 1, 2, 3, 4$. With identical weights, $\mathbf{X}\left(\tilde{\lambda}\right)$ is the projection of $\mathbf{T}$ to $\mathbf{X}(\lambda)$. If our assertion, that the response surface and the approximating plane are close, is valid then imposing potentials $\left\{X_1\left(\tilde{\lambda}\right), X_2\left(\tilde{\lambda}\right), X_3\left(\tilde{\lambda}\right), X_4\left(\tilde{\lambda}\right)\right\}$ on the master nodes will move the system close to the target state $\mathbf{T}$ [see Figures 1(b) and (c)].

## Results

We first analyze the planar approximation and its deviation from the response surface. The input data are the master node potentials for the unperturbed circuit and the four single knockout mutants, which are given in Table 1. $\mathbf{X}^{(n)}$, $n = 0, 1, 2, 3, 4$ can be used to construct the planar approximation to the response surface within the 4-simplex defined by the five points. Figure 3(a) shows the cross section $X_7(X_1, X_3)$, the points $\mathbf{X}^{(n)}$ ($n = 0, 1, 3$) denoted $\mathcal{P}_n$, and the planar approximation.

Next, we estimate the differences between the solution surface and the planar approximation within the simplex $\mathcal{P}_0\mathcal{P}_1\mathcal{P}_2\mathcal{P}_3\mathcal{P}_4$. Since a large set of data points is needed to span this 4-dimensional region with sufficiently high resolution, we restrict consideration to several subsets of the simplex. As an example, let us investigate the approximations on the boundary $\mathcal{P}_0\mathcal{P}_1\mathcal{P}_3$. A total of $25 \times 25$ grid points ($x_1, x_3$) are selected within the relevant domain and the response of the 16-node circuit is recorded when the potentials of nodes 1 and 3 are pre-set at these grid values. The average magnitude of the vector of sixteen node potentials ($\in \mathbb{R}^{16}$) on the surface is 5.19 Volts. The proximity of the approximations to the response surface need to be compared with this value. The planar approximation is constructed via Eq. (1) using the node potentials $\mathbf{X}^{(0)}$, $\mathbf{X}^{(1)}$ and $\mathbf{X}^{(3)}$. We evaluate the magnitude of the difference between the experimental data and the planar approximation at each grid point ($x_1, x_3$). The mean value of this magnitude within $\mathcal{P}_0\mathcal{P}_1\mathcal{P}_3$ is 152 mV.

Data on three additional points are required to construct the quadratic approximation. They are selected close to the midpoints of the sides of $\mathcal{P}_0\mathcal{P}_1\mathcal{P}_3$, and denoted $\mathcal{Q}_0$, $\mathcal{Q}_1$ and $\mathcal{Q}_3$ in Figure 3(b). The mean distance between the response surface and the quadratic approximation is found to be 24 mV. Thus, the quadratic approximation is significantly closer to the response surface than the plane. Similar results are found on other boundaries of the simplex $\mathcal{P}_0\mathcal{P}_1\mathcal{P}_2\mathcal{P}_3\mathcal{P}_4$.

Most natural networks are subjected to internal and external stochastic actions. For example, genetically identical cells can display variable levels of gene expression or even distinct phenotypes[23–25]. Next, we wish to address the role of stochasticity in approximating a response surface. Stochastic effects are added (synthetically) to node potentials since the level of noise in the electrical circuit is significantly smaller than in most natural systems. Specifically, the response surface is left unchanged and we add Gaussian noise of magnitude 5% to the components of $\mathbf{X}^{(0)}$, $\mathbf{X}^{(1)}$ and $\mathbf{X}^{(3)}$. Since the approximations to the response surface depend directly on $\mathbf{X}^{(n)}$, one expects the deviations from the response surface to change by a similar level. (This is in stark contrast to computations of network interactions, which require nonlinear inversions, generally amplifying the errors significantly.) We have computed the differences between the response surface and the approximations for 10,000 such cases. The addition of noise increases the mean deviation of the planar approximation from response surface $\mathcal{P}_0\mathcal{P}_1\mathcal{P}_3$ from 152 mV to 255 mV, and that of the quadratic approximation from 24 mV to 207 mV. When the level of noise is increased to 10%, the deviations of the planar and quadratic approximations increase to

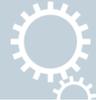

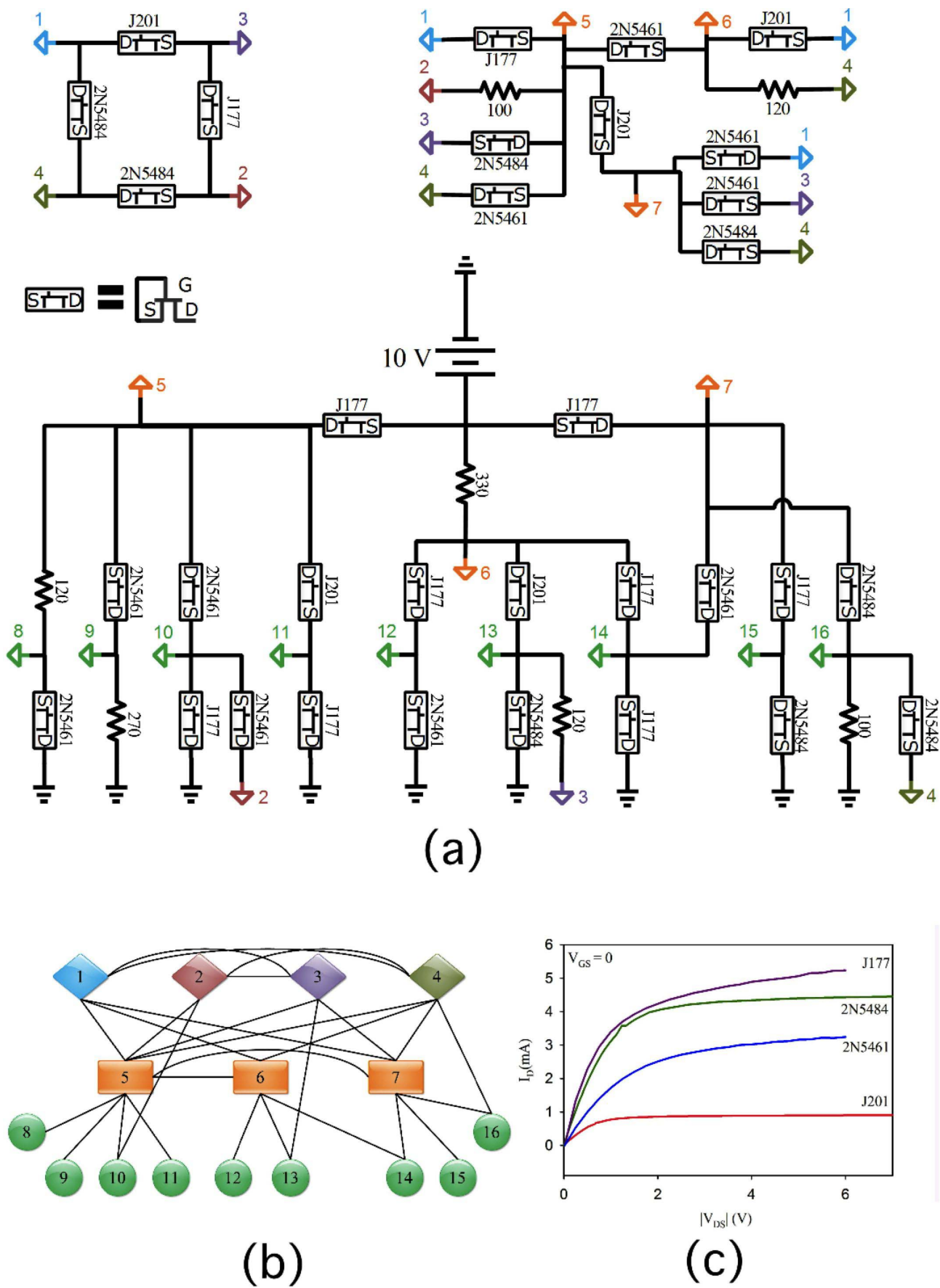

**Figure 2 | The electrical circuit used to test the network control algorithm.** (a) Circuit which models the network shown in (b). Rectangles in the circuit represent JFETs. Junctions corresponding to nodes on the upper levels of (b) are color-coded. (c) $I_D$ for the four JFETs used in the experiments.

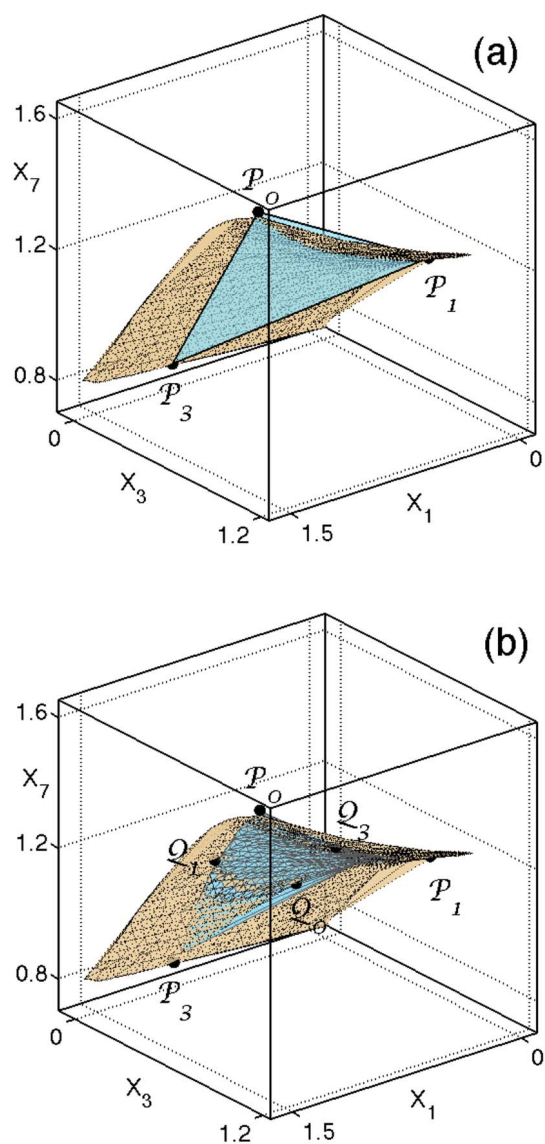

**Figure 3** | (a) Cross section of the response surface and the planar approximation. $\mathcal{P}_0$, $\mathcal{P}_1$, and $\mathcal{P}_3$ are the projections of the original state and the two single knockout mutants of nodes 1 and 3. (b) The cross section of the quadratic approximation that utilizes the two mutants (represented by $\mathcal{Q}_1$ and $\mathcal{Q}_3$) where $X_1$ and $X_3$ are individually set to half their values in the original system and one double mutant ($\mathcal{Q}_0$) where the values of both $X_1$ and $X_3$ are set to half their values in the original system. The quadratic surface is required to pass through $\mathcal{P}_0$, $\mathcal{P}_1$, $\mathcal{P}_3$, $\mathcal{Q}_0$, $\mathcal{Q}_1$ and $\mathcal{Q}_3$.

**Table 1** | Potentials, in Volts, of master nodes for the unperturbed circuit and for the four perturbations where one of the master nodes is grounded

|  | $X_1$ | $X_2$ | $X_3$ | $X_4$ |
|---|---|---|---|---|
| $\mathbf{X}^{(0)}$ | 1.614 | 1.461 | 1.148 | 1.847 |
| $\mathbf{X}^{(1)}$ | 0 | 0.988 | 0.665 | 1.513 |
| $\mathbf{X}^{(2)}$ | 0.903 | 0 | 0.403 | 1.381 |
| $\mathbf{X}^{(3)}$ | 0.995 | 0.902 | 0 | 1.427 |
| $\mathbf{X}^{(4)}$ | 0.428 | 0.352 | 0.372 | 0 |

450 mV and 430 mV respectively. These differences are small compared to the mean magnitude 5.19 V of node potentials on the response surface. We also note that, in this example, the advantage of using more refined approximations diminishes as the level of noise increases.

It is time and resource intensive to extract the entire response surface in examples like gene networks (where each point requires the sequencing of a genetically perturbed organism). Since we have already shown that a limited set of data points can provide a good approximation to a response surface, it is appropriate to derive error estimates using a small set of perturbations as well. For example, consider mutants where two master nodes are externally set to half their values in the unperturbed network. In our electrical circuit with four master nodes, there are six such mutants. Their mean prediction error using the planar approximations for the response surfaces is 38.4 mV. One can also consider double knockout mutants. Their mean error is 18.5 mV. The mean errors for triple and quadruple knockout mutants are 28.2 mV and 35.7 mV.

Next, we illustrate how the system can be moved close to the target. We note, first, that in examples such as gene networks, constraints on some nodes may be more important than those on others, an issue that can be resolved with appropriately defined weights, see Eq. (4). As an example, suppose we wish to move the network as close as possible to $T_6 = 3.0$, $T_8 = 0.9$, $T_{10} = 0.6$, $T_{12} = 2.3$, $T_{14} = 1.5$ and $T_{16} = 0.8$, the last three conditions being half as important. The relevant weights are $w_6 = w_8 = w_{10} = 1$; $w_{12} = w_{14} = w_{16} = 0.5$ (the remaining $w$'s = 0). Solving Eq. (4) gives $X_1 = 0.89$, $X_2 = 0.96$, $X_3 = 0.71$, and $X_4 = 1.44$. When these potentials are externally imposed on the master nodes, the electrical circuit reaches a state with $X_6 = 2.98$, $X_8 = 0.87$, $X_{10} = 0.56$, $X_{12} = 2.25$, $X_{14} = 1.49$ and $X_{16} = 0.68$; $U(\mathbf{T}; \mathbf{w})$ is $1.2 \times 10^{-2}$.

If the target $\mathbf{T}$ is far from the response surface, it is necessary to expand the set of master nodes (i.e., increase the dimensionality of the response surface) in order to reach sufficiently close to it. (For example, if we are required to increase $V_{16}$ 5-fold with no other changes, it is best to include node 16 in the master set.) Consider an example where the target state is $\mathbf{T}'$, which is identical to $\mathbf{T}$ except for $T'_{16} = 1.2$. We find that the closest point on $\mathbf{X}(\lambda)$ has $X_1 = 2.31$, $X_2 = 0.67$, $X_3 = 0.58$, and $X_4 = 1.51$. Imposing these node potentials on the circuit yields a state where $X_6 = 3.06$, $X_8 = 1.23$, $X_{10} = 0.78$, $X_{12} = 2.31$, $X_{14} = 1.73$ and $X_{16} = 0.75$; now $U(\mathbf{T}'; \mathbf{w})$ is 0.27. Closer approaches to $\mathbf{T}'$ cannot be made by controlling only the four master nodes.

The choice of additional master node(s) is made as follows: we compute $U(\mathbf{T}'; \mathbf{w})$ when each of $w_6$, $w_8$, $w_{10}$, $w_{12}$, $w_{14}$ and $w_{16}$ is individually set to zero. Small values for the corresponding $U(\mathbf{T}'; \mathbf{w})$ imply that the remaining nodes can be made to reach the point $\mathbf{T}'$ by altering the four master nodes, and hence that the selected node needs to be added to the master set. We find $U = 0.07$ when $w_6$ is set to zero, $U = 3.76 \times 10^{-4}$ when $w_8 = 0$, $U = 0.002$ when $w_{10} = 0$, $U = 0.08$ when $w_{12} = 0$, $U = 1.46 \times 10^{-5}$ when $w_{14} = 0$, and $U = 6.60 \times 10^{-5}$ when $w_{16} = 0$. Thus, one or more of nodes 8, 14, or 16 should be considered for inclusion in the master set. Using nodes 8 or 14 for the extension is experimentally impractical because some master node potentials require large changes from $\mathbf{X}^{(0)}$. The problem can be avoided by using node 16. The point on the (now 5-dimensional) response surface that minimizes $U$ is $X_1 = 1.79$, $X_2 = 0.76$, $X_3 = 0.35$, $X_4 = 1.33$, and $X_{16} = 1.20$. When these node voltages are imposed, the circuit reaches an equilibrium with $X_6 = 2.91$, $X_8 = 0.95$, $X_{10} = 0.56$, $X_{12} = 2.19$, and $X_{14} = 1.42$; $U(\mathbf{T}'; \mathbf{w})$ is $2.1 \times 10^{-2}$. Thus we have successfully moved the circuit close to $\mathbf{T}'$ by controlling nodes 1, 2, 3, 4, and 16.

We have shown how to compute the point on the approximating surface closest to the target and, if this point is not sufficiently close, how to expand the set of master nodes. The algorithm to reach the target can be initiated with a one (or a few) obvious master node(s). As the set is expanded, the dimensionality of the response surface increases, and states closer to the target are reached. As an example, consider convergence to a randomly selected target state. We select Node 3 as the first master node. Following the algorithm outlined above, we find that the next several master nodes are Nodes 6, 2, 16, 13, and 14. Observe that these are not necessarily regulatory nodes, shown in Figure 2(b). Figure 4 shows how the minimum distance $\delta$ between the target and the response surface, normalized by the distance $\delta_0$ between the original state and the target, decreases as the number of master nodes increases. In most cases, $\delta/\delta_0 \sim 20\%$ within 3–4 master nodes. Addition of noise does not change the results significantly.

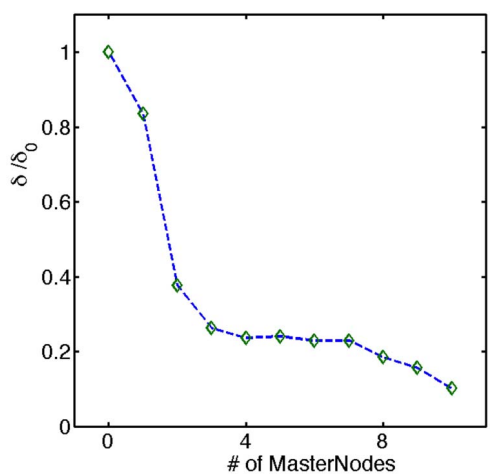

**Figure 4** | **The decay of the closest distance $\delta$ of the target to the response surface as the number of master nodes increases.** $\delta_0$ is the distance between the original state of the circuit and the target.

The circuit of Figure 2(a) only contains activating interactions (*i.e.*, $\partial V_i/\partial V_j > 0$ for each direct connection from node $i$ to node $j$). Hence, all direct and indirect (*i.e.*, with intermediary nodes) paths from node $i$ to node $j$ are activating. Consequently, the system is a *monotone network*[26], a class of networks whose solutions are highly robust[26,27]. Most natural networks contain both activating and inhibitory interactions, although they are conjectured to be *near-monotone*[26]. An obvious question that can be raised is whether the success of the response-surface-based control mechanism is related to the circuit being monotone.

To address this issue, we present an analysis of the network shown in Figure 5. It contains JFETs as well as Operational Amplifiers which are used to produce inhibitory interactions, *i.e.*, $\partial V_i/\partial V_j < 0$. The response surfaces of this network are found to be smooth as well. As seen in Figure 6, the plane passing through the original state and two mutants is found to be a good approximation to the surface. Importantly, it was possible to control this network to target states using the response-surface approach. Further details on the response surfaces and control of this network can be found in the Supplementary Materials. These results confirm that the efficacy of our control algorithm is not restricted to monotone networks.

## Discussion

Precise controllability is a major goal of network analyses[11]. A network can be controlled if an accurate model were available; unfortunately, such detailed information is rarely available for complex systems. Other approaches to control nonlinearly coupled systems require the network topology[13,14] or the ability to follow the evolution of a perturbed state experimentally[17]. Network topologies of many real systems are not available, and in examples like genetic networks, it is extremely challenging to follow the evolution experimentally. We have introduced an alternative approach for control that relies only on (the static) network responses to perturbations; typically, measurement of these responses is relatively straightforward.

The first step is to identify a preliminary set of master nodes to be used for control. In cases where the network topology is known, nodes of high out-degree can be selected for the purpose. When the topology is unknown, other information (*e.g.*, transcription factors in gene networks) must be used for the selection. The coarsest approximation to the response surface is a plane passing through the original state and the single knockout mutants; the input data for the algorithm are these states. Computation of more refined approximations to a response surface, such as a quadratic approximation, requires the states of additional mutants. Once an approximation is evaluated, the point on it that minimizes the weighted-squared-distance from the target state is computed. Our control strategy is to alter the master nodes to their values at this closest point.

Since the dimension of the response surface is significantly smaller than that of the state space, the nearest point on it may not be sufficiently close to the target. In such cases, it is necessary to expand the set of master nodes. We have shown that nodes to be included in the expansion can be deduced using suitable weighted distances. Following this process, control can be initiated with a small set of master nodes, which is expanded recursively to systematically move closer to the target. We emphasize that the final set of master nodes used for control will depend on the target state. *The approximations to response surfaces are all that are needed to implement control; we neither require the full set of nodes nor the network topology*. We note, however, that since our control algorithm relies only on stationary states, it cannot be used to perform dynamic feedback control.

In cases such as genetic networks, where the creation of mutants is expensive and/or time-consuming, data on double mutants can be used to estimate the deviation of the response surface from an approximation. In addition, the same data can be used to provide more refined approximations, such as quadratic approximations, to the response surface. We showed that typical quadratic approximations are significantly closer to the response surface than the corresponding planar approximations. However, the advantages of higher order approximations reduce as the level of stochasticity increases; the allocated resources may not offer comparable returns in noisy systems.

A significant advantage of our approach is that stochastic effects are not amplified in computing response surfaces. This is reasonable because the mutant data lie on the response surface, and inaccuracies in their measurements only change the response surface locally by a similar amount. These assertions were validated in the nonlinear electrical circuit. This robustness should be contrasted with computations of interactions between nodes which require nonlinear inversion; typically, errors are significantly amplified, especially in large systems.

We have tested our algorithm in networks whose sizes range from ~10 to ~50 nodes. Differences between the response surfaces and their planar or quadratic approximations are found to have similar magnitudes. Furthermore, the rate at which the target is approached [Figure 4] as a function of the number of master nodes exhibits similar behavior.

Our algorithm may need to be extended in the presence of bistability; *i.e.*, when a response surface is folded. Bistability in a slave node is of no concern since our control is implemented only on the master nodes. It may be relevant if the surface representing the state of a master node is folded as a function of the levels of other master nodes. One solution is to replace the bistable master node with another. Alternatively, one can use additional mutants to estimate the two branches of the folded surface.

We conclude with speculations on applications to a class of problems, namely genetic conversions of cellular states, where our control algorithm can potentially prove invaluable. Biological processes are governed by gene regulatory networks[18]. At present, neither the full set of nodes in many of these networks nor the interactions between the nodes are known with sufficient accuracy to model or control the systems. However, the states (*i.e.*, their genetic profiles) of the systems and of mutants are easily determined using microarrays or deep sequencing[21]. Furthermore, genetic networks are organized around nodes of high out-degree[4,18]. They include transcription factors (which control the production –and hence the expression levels– of many genes by binding to specific DNA sequences[28,29]) and/or micro RNAs (each of which down-regulates the levels of large collections of genes and transcription factors within cells[30]). Interestingly, transcription factors and microRNAs associated with a biological process are often known even in when the network topology is unknown or partially known. Some of these high out-degree nodes can be selected for the preliminary set of master nodes. The algorithm outlined above can then be used to expand the set of master nodes and to close in on the target. Note that the additional master nodes do not need to be of high out-degree.

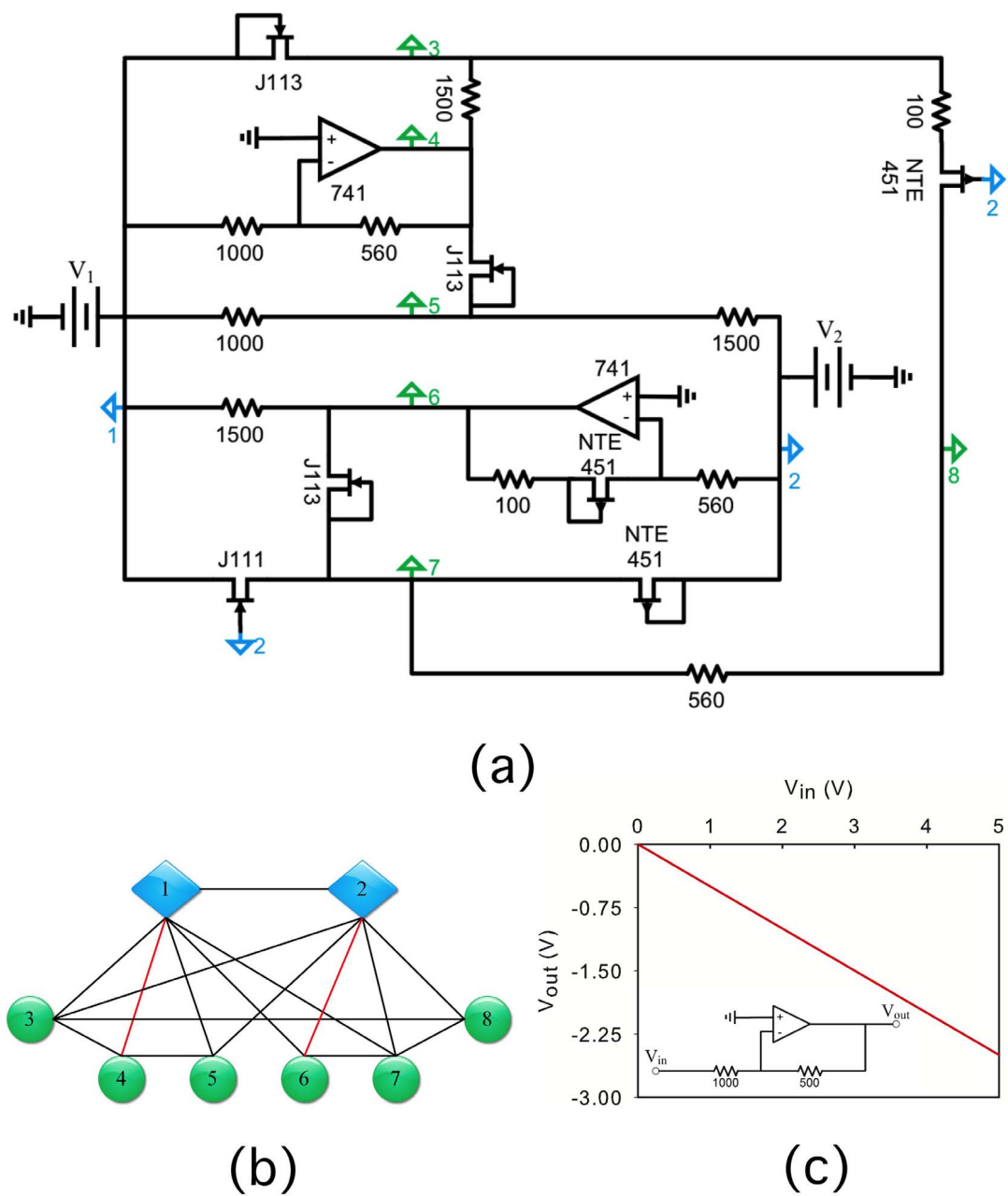

**Figure 5** | (a) Inhibitory interactions were produced with inverting amplifiers, each constructed from an Operational Amplifier (Op Amp) and a resistor, as in the interaction between node 1 and node 4 (and inset), and an Op Amp, a resistor, and a JFET for the effect of node 2 on node 6. (b) The network modeled by the circuit (a) where activating and inhibitory interactions are shown in black and red respectively. The network is not monotone. For example, the direct interaction 1 → 3 is activating while the indirect interaction 1 → 4 → 3 is inhibitory. (c) The output of an inverting amplifier shown in the inset as a function of the input.

One specific application may be in reprogramming fibroblasts (a class of connective tissue) to cardiomyocites (beating heart cells)[31,32]. It was demonstrated recently that the transformation can be achieved in culture using an assortment of three transcription factors Gata4, Mef2c, and Tbx5[31,32]. This set of genes was identified by studying the effects of adding all combinations of transcription factors associated with cardiac cell-fate[31]. Unfortunately, reprogramming efficiency and conversion to the mature cardiac phenotype remain low[32–34] and it is not clear if cardiac-like phenotype was maintained when the reprogrammed cells were transplanted into mouse hearts[32]. In order for these exciting findings to be used for heart repair, it is necessary to enhance the reprogramming efficiency and guarantee the robustness of the conversion. Two modifications may aid in these tasks: (1) Identifying the optimum levels at which the transcription factors are introduced into cells[32], and (2) Inclusion of additional genes of small or moderate out-degree. In the application of our algorithm, the states of fibroblasts and cardiomyocytes are the initial and target states. The preliminary set of master nodes can consist of Gata4, Mef2c, and Tbx5; their single knockout mutants can be used to derive the planar approximation to the three-dimensional response surface. The ratios in which the three genes should be introduced is evaluated from the closest point (on the plane) to the target. Additional nodes to be included in the master set (not necessarily high out-degree nodes) can be identified using weighed-distances.

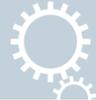

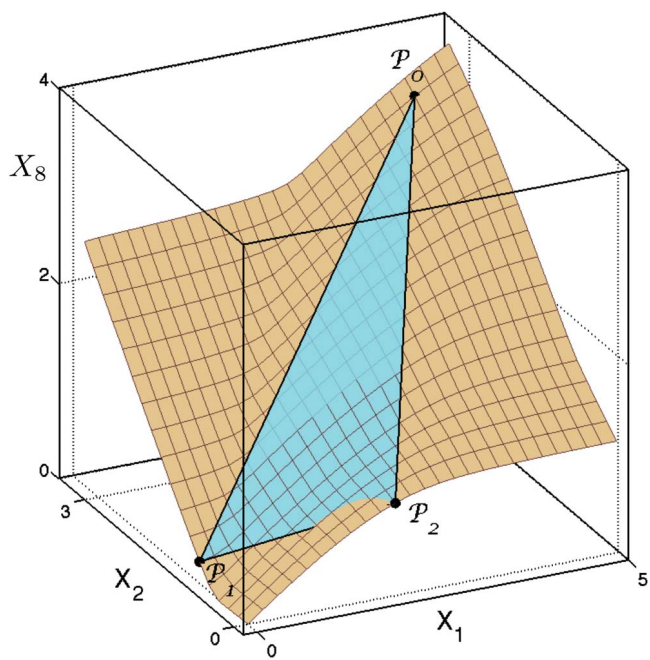

**Figure 6** | (a) Cross section of the response surface and the planar approximation. $\mathcal{P}_0$, $\mathcal{P}_1$, and $\mathcal{P}_2$ are the projections of the original state and the two single knockout mutants for the circuit of Figure 5.

### Acknowledgments
The authors would like to thank Drs. Preethi Gunaratne, Gregg Roman, Lars Seemann, Andréi Török, and Yuyi Xue for discussions and collaborations.

### Author contributions
J.S. and G.H.G. designed the experiment, conducted the analyses, and wrote the paper. J.S., F.M., A.M. and K.R. conducted the experiments.

### Additional information

**Supplementary information** accompanies this paper at http://www.nature.com/scientificreports

**Competing financial interests:** The authors declare no competing financial interests.

**How to cite this article:** Shulman, J., Malatino, F., Mo, A., Ryan, K. & Gunaratne, G.H. Controlling Networks of Nonlinearly-Coupled Nodes using Response Surfaces. *Sci. Rep.* **4**, 7574; DOI:10.1038/srep07574 (2014).